\documentstyle[12pt]{article}

\begin{document}
\begin{flushright}
\baselineskip=12pt

\end{flushright}

\begin{center}
\vglue 1.5cm
{\Large\bf Compactification and Supersymmetry Breaking in M-theory\\}
\vglue 2.0cm
{\Large  Tianjun Li}
\vglue 1cm
\begin{flushleft}
Department of Physics, University of Wisconsin, Madison, WI 53706
\end{flushleft}
\end{center}

\vglue 1.5cm
\begin{abstract}
Keeping N=1 supersymmetry in 4-dimension and in the
leading order, we disuss the various orbifold
compactifications of M-theory
 suggested by Horava and Witten
on $T^6/Z_3$, $T^6/Z_6$, $T^6/Z_{12}$, and the compactification
by keeping singlets under $SU(2)\times U(1)$
symmetry, then the compactification
on $S^1/Z_2$.
We also discuss the next to leading order K\"ahler potential,
superpotential, and gauge kinetic function in the $Z_{12}$ case.
In addition, we calculate the SUSY breaking soft terms and
find out that the universality of the scalar masses will be
violated, but the violation might be very small.

\end{abstract}

\vspace{0.5cm}
\begin{flushleft}
\baselineskip=12pt
January 1998\\
\end{flushleft}
\newpage
\setcounter{page}{1}
\pagestyle{plain}
\baselineskip=14pt

\section{Introduction}
Recently, M-theory on $S^1/Z_2$ suggested by Horava and
Witten~\cite{HW} has received
considerable attention. Its successes include as:
explanation of why Newton's constant appears to be so small
when the 4-dimensional grand unified gauge coupling $\alpha_{GUT}$
takes the experimental acceptable values~\cite{Witten},
gaugino condensation and
 supersymmetry breaking
~\cite{Horava, AQR, EDCJ, LT, LOW, CKM, NOY}, strong
CP problem~\cite{BD, KC},
threshhold scale and strong coupling effects~\cite{AQ,VK},
proton decay~\cite{EFN},
Casmir effect~\cite{MP},
compactification and low energy phenomenology consequences
~\cite{LLN, TIAN, LOD, DVN, BKL, BL, BJ}. In addition, this
approach may
open the door to M-theory model building
~\cite{JEDVN} and the testing M-theory models,
because there exists a crucial magnitude
difference in soft terms between
the M-theory and weakly coupled string theory, although their formulas
are similar.

As it is well known, a supergravity theory is specified by the
two functions, the K\"ahler function
\begin{eqnarray}
G &=& K +ln|W|^2 ~,~\,
\end{eqnarray}
where $K$ is the K\"ahler potential and $W$ is the superpotential,
and the gauge kinetic function $f$.
However, almost all the compactifications of this scenario
which have been
done is just the simplest case with the Hodge-Betti numbers of the
Calabi-Yau manifold $h_{(1,1)}=1, h_{(2, 1)} =0$  which
has just one modulus from the Calabi-Yau manifold.
In this paper,  we consider the various orbifold
compactifications
on $T^6/Z_3$, $T^6/Z_6$, $T^6/Z_{12}$ and the compactification
by retaining singlets under $SU(2)\times U(1)$ symmetry,
corresponding to the Hodge-Betti
numbers  are $h_{(1,1)}=9, 5, 3, 2$
 and $h_{(2, 1)}=0$ respectively, these models will have
9, 5, 3, 2 families respectively.
The compactification should be done from eleven-dimension
to five-dimension, then to the four-dimension, at leading order
($\kappa^{2/3}$); the K\"ahler potential, superpotential and gauge
kinetic function should be the same as previous~\cite{SFKP}
but the defintions of
the moduli are different, as pointed out in Ref~\cite{TIAN}.
The interesting
point is that all above models are
no-scale supergravity models~\cite{no-scale}.
To the higher order,
we should include the M-theory correction.
The 4-dimensional effective action of the M-theory can be expanded
in powers of the two kinds of dimensionless variables~\cite{CKM}:
\begin{eqnarray}
\epsilon_1 &=& \kappa^{2/3} \pi \rho /V^{2/3} ~,~\,
\end{eqnarray}
and
\begin{eqnarray}
\epsilon_2 &=& \kappa^{2/3} /(\pi \rho V^{2/3}) ~.~\,
\end{eqnarray}
For example, in the model with just two moduli S and T,
to the order of $\kappa^{4/3}$, the relevent
variable is $\epsilon_1$ ( or $ReT/ReS$ )
defined above~\cite{NOY, LOD}. In other words, when we choose
the metric as follows:
\begin{eqnarray}
g_{\mu \nu}^{(11)}=e^{-\gamma} e^{- 2\sigma } g_{\mu \nu}^{(4)} ~;~
g_{11, 11}^{(11)}=e^{2\gamma} e^{-2\sigma} ~;~
g_{m n}^{(11)}= e^{\sigma }g_{ m n} ~,~ \,
\end{eqnarray}
the expansion variable is $\kappa^{2/3} e^{\gamma -3\sigma}$.
The theory has more than two variables when one consider more
moduli~\cite{BL, BJ}.
Here, we consider in detail the model with $Z_{12}$ symmetry for detail.
Although the original Lagrangian suggested by Horava and
Witten is generally constructed up to the terms at
order of $\kappa^{2/3}$,  we write down
 K\"ahler potential,
superpotential and gauge kinetic function at the
order of $\kappa^{4/3}$
(the detailed calculation will be appeared elsewhere~\cite{BL}.),
which might be the part of
M-theory correction. The expansion variables now are:
\begin{eqnarray}
\epsilon_3 &=& \kappa^{2/3}
e^{\gamma -( \sigma_1 + 4\sigma_2 + 4\sigma_3)/3} \,
\end{eqnarray}
\begin{eqnarray}
\epsilon_4 &=& \kappa^{2/3}
e^{\gamma -(4\sigma_1 + \sigma_2 + 4 \sigma_3)/3} \,
\end{eqnarray}
\begin{eqnarray}
\epsilon_5 &=& \kappa^{2/3}
e^{\gamma -(4\sigma_1 + 4\sigma_2 +  \sigma_3)/3} \,
\end{eqnarray}
where of $\gamma $ and $\sigma_i$
are defined in the following sections. In addition, if we did not require
$Z_{12}$ symmetry but had four moduli: $S$, $T_1$, $T_2$ and $T_3$ in
the 4-dimension, there will be 6 expansion vaiables~\cite{BJ}.
Furthermore, we discuss
the non-pertubative
superpotential related to $S$ and $T_i$. Finally,
we calculate the SUSY-breaking soft terms under the general assumption
that SUSY is spontaneouly broken by the auxiliary components of the bulk
modui superfields in the model. We also discuss dilaton-induced
SUSY-breaking, because it is simple and it is special in several respects
in previous considerations. We emphasize that, although the
K\"ahler potential, superpotential and gauge kinetic function might
have similar forms as before, the magnitude is much different after
we consider the next order correction, so there
can be sizable differences in the weakly coupled string
and M-theory. These results might be helpful for
future M-theory model building and phenomenology analysis, and one
might find the differences in the future accelerator experiment.

Of course, if $\int w \wedge tr( R \wedge R ) $ is zero, then the next 
leading order correction will be zero and there may be no strong  
coupling
in the hidden sector. It may be also possible to break SUSY by
Scherk-Scherk mechanism~\cite{AQR, EDCJ}
because of the $Z_2^{HW}$ symmetry in this scenario.
However, the length of the eleven-dimension could not take very
small values because we should use the acceptable $\alpha_{GUT}$,
$m_{GUT}$ and $m_{pl}$. The possible range of $\pi \rho $ is from
$10^{13} ~GeV$ to $10^{15} ~GeV$ ~\cite{BKL} when we use
$m_{GUT}= 2.0 \times 10^{16} ~ GeV$ estimated by extrapolating from
the measurements at LEP and elsewhere.
If so, the results
discussed in Section 2 will not be changed in the next leading
order ($\kappa^{4/3}$).

Obviously, the discussion in this manuscript is not the general case,
the general cases of  dilaton/moduli induced SUSY-breaking and the
M-theory corrections to previous string no-scale
supergravity~\cite{JLDVN} are under investigation.

This manuscript is organized as follows: in  Section 2, we
calculate the K\"ahler potential, superpotential and gauge kinetic
function at order of $\kappa^{2/3}$. We generalize the result
to the order of $\kappa^{4/3}$ in $Z_{12}$ symmetry case
in Section 3. In Section 4, we
discuss its the gaugino condensation and corresponding superpotential. 
In Section 5, we discuss its SUSY-breaking soft terms. Our conclusions
are given in section 6.
\section{SUGRA of the order $\kappa^{2/3}$}
The four-dimensional effective action in M-theory scenario has been 
calculated in the simplest case with just S and one T fields.
As previously, the strategy consists in
a dimensional reduction of the 11-dimensional Lagrangian on $S^1/Z_2$.
A consistent truncation is performed by keeping only singlets
or invariance with respect
to some suitable group, isomorphic to a
subgroup of the rotation group of the
internal manifold and then compactified on $S^1/Z_2$ to obtain a
N=1 supersymmetry theory. For example, the 4-dimensional effective
actions~\cite{TIAN}
are obtained by keeping only the singlets under the action of
$SU(3)_D=SU(3)_1 + SU(3)_2$, in which $SU(3)_1$ is a subgroup of  
the rotation
group SO(6) or SU(4) of the internal coodinates $X_I$, where I=5,  
..., 10 in
11-dimension, and $SU(3)_2\times E_6 \supset E_8$ where $E_8$ is  
the gauge group in the observable sector or boundary.

Using methods in Ref.~\cite{EDCJ, TIAN, SFKP},
we generalize this strategy to obtain
other N=1, D=4 effective actions in M-theory. With the boundary gauge
group $G=E_8\times E_8'$, one can obtain a wide range of the N=1
supergravity models with families in the 27 representation of
$E_6 \supset E_8$. What has been already done is the one family
model which keeps only the $SU(3)_D$ singlets~\cite{TIAN} or just keeps
two moduli: $S$ and $T$~\cite{LOD, NOY}; one can compactify
 on the orbifolds by restriction to the subgroup of SU(3), and then
obtain other models. The N=1 ``maximal model" with 9 families is obtained
 by choosing
$Z_3$ symmetry.
The lower families model can be obtained by a suitable
truncation of the ``maximal model". As pointed in Ref. ~\cite{TIAN}, 
to order $\kappa^{2/3}$, the result is the same as
previous result except the
definition of the dilaton and the moduli. After some exercise,
one can obtain the following 4-dimensional K\"ahler potential,
superpotential and the gauge kinetic function.

(I) $Z_3$ symmetry. This is the center $Z_3$ of the $SU(3)_D$. The
model contains nine 27 of families which are triplets under the  
horizontal
gauge SU(3) symmetry. The massless fields in 5-dimension
are the gravitational multiplet, the universal hypermultiplet and the 
eight vector multiplets, corresponding to the Hodge-Betti
number is $h_{(1,1)}=9$ and
$h_{(2, 1)}=0$\footnote{Generally, in 5-dimension, the theory contains
one gravitational multiplet, ($h_{1, 1} - 1$) vector multiplets,
and ($h_{2, 1} + 1$) hypermultiplets.}; and the observable gauge group
is $SU(3)\times E_6$ ( the hidden sector gauge group is $E_8$).
In the Einstein frame after rescaling, the $Z_3$ invariant metric tensor
in 5-dimension is:
\begin{eqnarray}
g_{\mu \nu}^{(11)}= G^{-1/3} g_{\mu \nu}^{(5)} ~;~
g_{11, 11}^{(11)}= G^{-1/3} g_{11, 11}^{(5)} ~;~
g_{i\bar j}^{(11)}= g_{i\bar j} ~,~ \,
\end{eqnarray}
and in 4-dimension is:
\begin{eqnarray}
g_{\mu \nu}^{(5)}=e^{-\gamma} g_{\mu \nu}^{(4)} ~;~
g_{11, 11}^{(5)}=e^{2\gamma} ~,~ \,
\end{eqnarray}
where the G is the determinant of the metric in the compact  
6-dimension space.
The shape of the orbifold is described by the nine parameters  
$g_{i\bar j}$
where i, j =1, 2, 3. The massless modes of the three form in
4-dimension is $C_{5\mu \nu}$ and $C_{5 i \bar j}$.
The K\"ahler potential
for 4-dimensional SUGRA is
\begin{eqnarray}
K &=& -\ln\,[S+\bar S]-\ln\, det[T_{i \bar j} +\bar T_{i \bar j}-2
C_i^{n a} C_{\bar j}^{\bar n \bar a}]  ~,~ \,
\end{eqnarray}
where
\begin{eqnarray}
S=G^{1/2}+ i 24 \sqrt 2 D ~,~\,
\end{eqnarray}
and
\begin{eqnarray}
T_{i \bar j}= e^{\gamma}G^{-1/6}g_{i\bar j} - i 6 \sqrt 2 B_{i\bar j}
+ C_i^{n a} C_{\bar j}^{\bar n \bar a} ~,~\,
\end{eqnarray}
with $ {1\over {4!}} G G_{\mu \nu \rho 5} =
\epsilon_{\mu \nu \rho \sigma}
\partial^{\sigma} D$, and $C_{5 i \bar j}= i B_{i\bar j} $.
The gauge kinetic function for the observable and hidden sector is
\footnote{One may have a normalization factor f~\cite{TIAN},
but here, we
put it as 1; otherwise one should write:
\begin{eqnarray}
f_{\alpha \beta} &=& f S\, \delta_{\alpha \beta} ~.~\,
\end{eqnarray}
}
\begin{eqnarray}
f_{\alpha \beta} &=&  S\, \delta_{\alpha \beta} ~,~ \,
\end{eqnarray}
and the superpotential $W$,
\begin{eqnarray}
 W=  c~ g_c ~ \epsilon^{i j k} ~
(\epsilon_{m n l} ~ d_{a b c}) C_i^{m a}
 C_j^{n b} C_k^{l c} ~,~\,
\end{eqnarray}
where c is just a constant and $g_c ={{ 2\pi \rho \lambda^2}
\over\displaystyle {\kappa^2 }}$.
And $\lambda^2=2\pi (4 \pi \kappa^2)^{2/3}$. In addition,
$\epsilon_{i j k}$ is the antisymmetric tensor of SU(3),
$d_{a b c}$ is the symmetric tensor of the 27 of the $E_6$,
a $\in$ 27 of $E_6$.

The ``maximal" nine family K\"ahler manifold defined from above
is
\begin{eqnarray}
{{SU(1, 1)}\over\displaystyle {U(1)}}\times
{{SU(3, 3 + 3 n)}\over\displaystyle
{ SU(3)\times SU(3 + 3 n)\times U(1)}} ~,~\,
\end{eqnarray}
where n=27.

(II) $Z_6$ symmetry. $Z_6$ symmetry is generated by the 3 $\times$ 3
matrix:
\begin{eqnarray}
M&=&diag(-e^{i2\pi /3}, -e^{i2\pi /3}, e^{i2\pi /3})\,
\end{eqnarray}
The $Z_6$ invariance keeps only singlets and triplets under
$SU(2)_D \supset SU(3)_D$. The
model contains five 27 families. In other words, the massless fields 
in 5-dimension
are gravitational multiplet, the universal hypermultiplet and
four vector multiplets, corresponding to the Hodge-Betti
number is $h_{(1,1)}=5$ and $h_{(2, 1)}=0$, and the observable  
gauge group
is $SU(2)\times U(1) \times E_6$ ( the hidden sector gauge group is  
$E_8$).
In the Einstein frame after rescaling, the $Z_6$ invariant metric tensor
in 4-dimension is
\begin{eqnarray}
g_{\mu \nu}^{(11)}=e^{-\gamma} G^{-1/3} g_{\mu \nu}^{(4)} ~;~
g_{11, 11}^{(11)}=e^{2\gamma} G^{-1/3} ~,~\,
\end{eqnarray}
\begin{eqnarray}
g_{i\bar j}^{(11)}= g_{i\bar j} ~,~
g_{3\bar l}^{(11)}= g_{3\bar 3} \delta_{3\bar l} ~,~\,
\end{eqnarray}
where $i, j = 1, 2$ and $l = 1, 2, 3$;
G is the derminant of the metric in the compact 6-dimension space.
The shape of the orbifold is described by the five parameters  
$g_{i\bar j}$
where i, j =1, 2 and $g_{3\bar 3}$.
The massless modes of the three form in 4-dimension is $C_{5\mu \nu}$ 
and $C_{5 i \bar j}$. The K\"ahler potential
for 4-dimensional SUGRA is
\begin{eqnarray}
K &=& -\ln\,[S+\bar S]- \ln\, det[T_{i \bar j} +\bar T_{i \bar j}-2 
C_i^{m a} C_{\bar j}^{\bar m \bar a}]
-\ln\,[T_3 +\bar T_3 -2 C^a C^{\bar a}] ~,~ \,
\end{eqnarray}
where
\begin{eqnarray}
S=G^{1/2}+i 24 \sqrt 2 D ~,~ \,
\end{eqnarray}
and
\begin{eqnarray}
T_{i \bar j}= e^{\gamma}G^{-1/6}g_{i\bar j} - i
6 \sqrt 2 B_{i\bar j}
+ C_i^{m a} C_{\bar j}^{\bar m \bar a} ~,~\,
\end{eqnarray}
\begin{eqnarray}
T_3= e^{\gamma}G^{-1/6}g_{3\bar 3} -  i
6 \sqrt 2 B_3
+ C^a C^{\bar a} ~,~\,
\end{eqnarray}
where ${1\over {4!}} G G_{\mu \nu \rho 5}
= \epsilon_{\mu \nu \rho \sigma}
\partial^{\sigma} D$, $C_{5 i \bar j} =i B_{i\bar j}$
and $C_{5 3 \bar 3}= i B_3$;
here, $i, j, m = 1, 2$.
The gauge kinetic function for the observable and hidden sector is:
\begin{eqnarray}
f_{\alpha \beta} &=&  S\, \delta_{\alpha \beta} ~,~ \,
\end{eqnarray}
and the superpotential $W$ is
\begin{eqnarray}
 W= c ~ g_c ~ \epsilon^{i j} ~
(\epsilon_{m n} ~ d_{a b c}) C_i^{m a}
 C_j^{n b} C^c ~,~\,
\end{eqnarray}
where $\epsilon_{ i j}$ is the antisymmetric tensor of SU(2).

The five family K\"ahler manifold defined from above
is
\begin{eqnarray}
{{SU(1, 1)}\over\displaystyle {U(1)}}\times
{{SU(2, 2 + 2 n)}\over\displaystyle
{ SU(2)\times SU(2 + 2 n)\times U(1)}}
\times{{SU(1, 1+n)}\over\displaystyle
{SU(1+n)\times U(1)}} ~,~\,
\end{eqnarray}
where n=27.

(III) $Z_{12}$ symmetry. $Z_{12}$ symmetry is generated by the 3  
$\times$ 3
matrix:
\begin{eqnarray}
M&=&diag(ie^{i2\pi /3}, -ie^{i2\pi /3}, e^{i2\pi /3}) ~,~\,
\end{eqnarray}
The
model contains three 27 families. In other words, the massless fields 
in 5-dimension
are gravitational multiplet, the universal hypermultiplet and
two vector multiplets, corresponding to the Hodge-Betti
number is $h_{(1,1)}=3$ and $h_{(2, 1)}=0$, and the observable  
gauge group
is $U(1)\times U(1) \times E_6$ ( the hidden sector gauge group is  
$E_8$).
In the Einstein frame after rescaling, the $Z_6$ invariant metric tensor
is:
\begin{eqnarray}
g_{\mu \nu}^{(11)}=e^{-\gamma -2 (\sigma_1 + \sigma_2
+ \sigma_3) / 3} g_{\mu \nu}^{(4)} ~;~
g_{11, 11}^{(11)}=e^{2\gamma - 2 (\sigma_1 + \sigma_2
+ \sigma_3) / 3}~;~
g_{i\bar j}^{(11)}= e^{\sigma_i} \delta_{i\bar j} ~,~\,
\end{eqnarray}
where $i, j = 1, 2, 3$.
The shape of the orbifold is described by the three parameters:
$\sigma_i$.
The massless modes of the three form in 4-dimension is $C_{5 \mu \nu}$ 
and $C_{5 i \bar i}$. The K\"ahler potential
for 4-dimensional SUGRA is:
\begin{eqnarray}
K &=& -\ln\,[S+\bar S]- \sum_{i=1}^3
\ln\,[T_i +\bar T_i -2 C_1^a C_1^{\bar a}] ~,~ \,
\end{eqnarray}
where
\begin{eqnarray}
S=G^{1/2}+i 24 \sqrt 2 D ~,~\,
\end{eqnarray}
and
\begin{eqnarray}
T_i= e^{\gamma - (\sigma_1 + \sigma_2
+ \sigma_3) / 3 + \sigma_i }
- i 6 \sqrt 2 B_i
+ C_i^a C_{\bar i}^{\bar a} ~,~\,
\end{eqnarray}
where ${1\over {4!}} e^{2 (\sigma_1 + \sigma_2
+ \sigma_3)} G_{\mu \nu \rho 5} = \epsilon_{\mu \nu \rho \sigma}
\partial^{\sigma} D$, $C_{5 i \bar i} = i B_i$.
The gauge kinetic function for the observable and hidden sector is:
\begin{eqnarray}
f_{\alpha \beta} &=&  S\, \delta_{\alpha \beta} \,
\end{eqnarray}
and the superpotential $W$ is
\begin{eqnarray}
 W= c ~ g_c ~ d_{a b c} ~  C_1^a
 C_2^b C_3^c ~.~\,
\end{eqnarray}
The  three family K\"ahler manifold defined from above
is
\begin{eqnarray}
{{SU(1, 1)}\over\displaystyle {U(1)}}\times
({{SU(1, 1+n)}\over\displaystyle
{SU(1+n)\times U(1)}})^3 ~,~\,
\end{eqnarray}
where n=27. Also note that, three families have charges ( 1, 1 ),
( -1, 1 ), ( 0, -2 ) under two U(1), respectively.

(IV) $SU(2)\times U(1)$ symmetry.This
model contains two 27 families. In other words, the massless fields 
in 5-dimension
are gravitational multiplet, the universal hypermultiplet and the
one vector multiplet, corresponding to the Hodge-Betti
number is $h_{(1,1)}=2$ and $h_{(2, 1)}=0$, and the observable  
gauge group
is $U(1) \times E_6$ ( the hidden sector gauge group is $E_8$).
In the Einstein frame after rescaling, the $SU(2)\times U(1)$
invariant metric tensor is:
\begin{eqnarray}
g_{\mu \nu}^{(11)}=e^{-\gamma - (4 \sigma + 2 \sigma_3 ) / 3}
g_{\mu \nu}^{(4)} ~;~
g_{11, 11}^{(11)}=e^{2\gamma - (4 \sigma
+ 2 \sigma_3) / 3} ~,~\,
\end{eqnarray}
\begin{eqnarray}
g_{i\bar j}^{(11)}= e^{\sigma} \delta_{i\bar j}
~;~
g_{3 \bar l}^{(11)}= e^{\sigma_3} \delta_{3 \bar l} ~,~\,
\end{eqnarray}
where $i, j = 1, 2$ and $l = 1, 2, 3$.
The shape of the orbifold is described by the two parameters:
$\sigma$ and
$\sigma_3$.
The massless modes of the three form in 4-dimension is $C_{5 \mu \nu}$ 
and $C_{5 i \bar i}$. The K\"ahler potential
for 4-dimensional SUGRA is
\begin{eqnarray}
K &=& -\ln\,[S+\bar S]- 2 \ln\,[T+\bar T-2 C_T^a C_T^{\bar a}] -
\ln\,[T_3 +\bar T_3 -2 C_3^a C_3^{\bar a}] ~,~ \,
\end{eqnarray}
where
\begin{eqnarray}
S=G^{1/2}+i 24\sqrt 2 D ~,~\,
\end{eqnarray}
and
\begin{eqnarray}
T= e^{\gamma - (2 \sigma
+ \sigma_3) / 3 + \sigma } - i 6 \sqrt 2 B
+ C_T^a C_T^{\bar a} ~,~\,
\end{eqnarray}
\begin{eqnarray}
T_3= e^{\gamma - (2 \sigma +
\sigma_3) / 3 + \sigma_3 } - i 6 \sqrt 2 B_3
+ C_3^a C_3^{\bar a} ~,~\,
\end{eqnarray}
where ${1\over {4!}} e^{4 \sigma
+ 2 \sigma_3} G_{\mu \nu \rho 5} = \epsilon_{\mu \nu \rho \sigma}
\partial^{\sigma} D$, $C_{5 i \bar i}= i B$ where i= 1, 2
and $C_{5 3 \bar 3}= i B_3$.
The gauge kinetic function for the observable and hidden sector is
\begin{eqnarray}
f_{\alpha \beta} &=&  S\, \delta_{\alpha \beta} \,
\end{eqnarray}
and the superpotential $W$ is
\begin{eqnarray}
 W= c ~ g_c ~ d_{a b c} ~ C_T^a
 C_T^b C_3^c ~.~\,
\end{eqnarray}
The  two family K\"ahler manifold defined from above
is
\begin{eqnarray}
{{SU(1, 1)}\over\displaystyle {U(1)}}\times
({{SU(1, 1+n)}\over\displaystyle
{SU(1+n)\times U(1)}})^2\,
\end{eqnarray}
where n=27. In addition,  $C_T$ and $C_3$ have +1 and -2 U(1) charges, 
respectively.
\section{Some SUGRA Terms of the order $\kappa^{4/3}$}
One can not compute the full effective action to the order of
$\kappa^{4/3}$ because the original 11-dimensional theory is  
generally constructed up to terms of the order of
$\kappa^{4/3}$. Here, we just consider order $\kappa^{4/3}$
terms involving gauge or gauge matter fields, and do not consider
higher derivative terms and terms of higher mass dimension~\cite{LOD}.

To order $\kappa^{4/3}$, we should consider the deformed Calabi-Yau
manifold or orbifold in order to have N=1 supersymmetry
in 4-dimension. Moreover,
the effective action of M-theory can be annalyzed in the simple
case by expanding it in
one dimensionless variable~\cite{CKM, NOY, LOD}:
\begin{eqnarray}
\epsilon_1 &=& \kappa^{2/3} \pi \rho /V^{2/3} \,
\end{eqnarray}
To calculate the 4-dimensional effective action, one first
expands the 11-dimensional action in powers of $\kappa^{2/3}$
to obtain the compactification solution which is expanded
in $\epsilon_1$. The subsequent
Kaluza-Klein reduction of the 11-dimensional action  for this
compactification solution will lead to the desired 4-dimensional
effective action. The compactification which has been down is
following above line.

In the $Z_3$ symmetry, the M-theory correction will be
very complicated, and one should also keep in mind
that there are only
three families in the low energy phenomenology. Therefore,
in the following
parts, we will only pay attention to the $Z_{12}$ case. In this
case, the physical volume of the Calabi-Yau manifold is
$V e^{\sigma_1 +\sigma_2 + \sigma_3}$, so we will have more
expansion variables. The technical details will be given
elsewhere~\cite{BL}; here
we just want to consider the general picture. We write down
 the K\"ahler potential, superpotential and
gauge kinetic function. Notice that we can also write the
$Z_{12}$ invariant metric as the following:
\begin{eqnarray}
g_{\mu \nu}^{(11)}=e^{-\gamma -2 (\sigma_1 + \sigma_2
+ \sigma_3) / 3} g_{\mu \nu}^{(4)} ~;~
g_{11, 11}^{(11)}=e^{2\gamma - 2 (\sigma_1 + \sigma_2
+ \sigma_3) / 3}~;~
g_{i\bar j}^{(11)}= e^{\sigma_i} g_{i\bar i}^{I} \delta_{i\bar j} ~,~\,
\end{eqnarray}
where i, j = 1, 2, 3. $g_{i \bar i}^{I}$ is a function of the coordinates
in 6-dimensional compact space which is also satisfy $Z_{12}$ invariance
and V in above paragraph can be expressed as
$V= \int d^6 x \sqrt{g^{I}}$.

In the Einstein frame after rescaling,  the K\"ahler potential
for 4-dimensional SUGRA is
\begin{eqnarray}
K &=& -\ln\,[S+\bar S]- \sum_{i=1}^3
\ln\,[T_i +\bar T_i -2 C_{ia}^* C_i^a]
\nonumber\\&&
+  {2\over 3} {1\over {S+\bar S}}
(\sum_{j=1}^3 \alpha_j (T_j + \bar T_j))
\sum_{i=1}^3({{C_{ia}^* C_i^a}\over\displaystyle
{T_i +\bar T_i}})
~,~ \,
\end{eqnarray}
where
\begin{eqnarray}
S=G^{1/2}+i 24 \sqrt 2 D  ~,~\,
\end{eqnarray}
and
\begin{eqnarray}
T_i= e^{\gamma - (\sigma_1 + \sigma_2
+ \sigma_3) / 3 + \sigma_i } - i 6 \sqrt 2 B_i
+ C_{ia}^* C_i^a ~,~\,
\end{eqnarray}
The gauge kinetic functions for the observable and hidden sector
 sector are\footnote{Exactly speaking, we should write the gauge kinetic
funtion as:
\begin{eqnarray}
f^o_{\alpha \beta} &=&  (S + \sum_{i=1}^3
\alpha_i (T_i - C_{ia}^* C_i^a))
 \, \delta_{\alpha \beta} ~,~\,
\end{eqnarray}
\begin{eqnarray}
f^h_{\alpha \beta} &=& (S - \sum_{i=1}^3
\alpha_i (T_i - C_{ia}^* C_i^a))
 \, \delta_{\alpha \beta} ~.~\,
\end{eqnarray}
But, we always consider $ < C_i^a > = 0 $, so, we write as above
forms. }
\begin{eqnarray}
f^o_{\alpha \beta} &=&  (S + \alpha_1 T_1 +
\alpha_2 T_2 + \alpha_3 T_3))
 \, \delta_{\alpha \beta} ~,~\,
\end{eqnarray}
\begin{eqnarray}
f^h_{\alpha \beta} &=&  (S - \alpha_1 T_1 -
\alpha_2 T_2 - \alpha_3 T_3))
 \, \delta_{\alpha \beta} ~,~\,
\end{eqnarray}
and the superpotential $W$ is
\begin{eqnarray}
 W=  c ~g_c ~ d_{a b c} ~ C_1^a
 C_2^b C_3^c ~.~\,
\end{eqnarray}
The three expansion variables are
\begin{eqnarray}
\epsilon_3 &=& \kappa^{2/3}
e^{\gamma -( \sigma_1 + 4\sigma_2 + 4\sigma_3)/3} ~,~\,
\end{eqnarray}
\begin{eqnarray}
\epsilon_4 &=& \kappa^{2/3}
e^{\gamma -(4\sigma_1 + \sigma_2 + 4 \sigma_3)/3} ~,~\,
\end{eqnarray}
\begin{eqnarray}
\epsilon_5 &=& \kappa^{2/3}
e^{\gamma -(4\sigma_1 + 4\sigma_2 +  \sigma_3)/3} ~.~\,
\end{eqnarray}
Of course, if we chose the following metric:
\begin{eqnarray}
g_{\mu \nu}^{(11)}=e^{-\gamma -2 (\sigma_1 + \sigma_2
+ \sigma_3) / 3} g_{\mu \nu}^{(4)} ~;~
g_{11, 11}^{(11)}=e^{2\gamma - 2 (\sigma_1 + \sigma_2
+ \sigma_3) / 3}~;~
g_{i\bar j}^{(11)}= e^{(\sigma_i + \sigma_j)/2} g_{i \bar j}^{CY} \,
\end{eqnarray}
we will also have a model with moduli: $S$, $T_1$, $T_2$ and $T_3$.
However, in this case, the K\"ahler potential and gauge kinetic
function will be a little more complicated and we will have 6 expansion
variables~\cite{BJ}.

\section{Effective Superpotential by the Gaugino Condensation}
For the case of two moduli $S$ and $T$ in M-theory,
 the superpotential
and gaugino condensation were
analyzed~\cite{LT, LOW}. The
arguments should not be changed when we consider  $Z_{12}$
symmetry. The superpotential in M-theory
is similar to that in the
weakly coupled string theory because to the leading order and
next to the leading order, their results are similar.
In short,
 one can write the following superpotential as
previously~\cite{MRNE}
\footnote{In order to
keep the soft breaking
terms not too large, one
might consider $c = 0$~\cite{LOW}, i.e.,
the following superpotential:
\begin{eqnarray}
W=m_{11}~ h~ exp(-{3 \over\displaystyle {2 b_0 g_h^2}} ) \,
\end{eqnarray}
}:
\begin{eqnarray}
W=m_{11}(c + h ~exp(-{3 \over\displaystyle {2 b_0 g_h^2}} ))\,
\end{eqnarray}
Where $m_{11}$ is the 11-dimensional Planck scale and
$m_{11}=\kappa^{2/9}$,  $b_0$ is the coefficient of the one-loop
$\beta $ function in the hidden sector, and h and c
are constants.
Notice 11-dimensional Lagrangian of the Yang-Mills Fields in
the hidden sector is
\begin{eqnarray}
L_B&=&
{1\over\displaystyle 2\pi (4\pi \kappa^2)^{2\over 3}}
\int_{M^{10}}d^{10}x\sqrt g {1\over 4}F_{AB}^aF^{aAB}\,
\end{eqnarray}
The 4-dimensional gauge coupling in the hidden sector is
\begin{eqnarray}
g_h^{-2}&=&{V_p^h \over\displaystyle
{ 2\pi (4\pi \kappa^2)^{2\over 3}}} \,
\end{eqnarray}
Where $V_p^h$ is the physical 6-dimensional volume in the hidden
sector and
$V_p^h = V f^h$. Assuming $V^{-1/6}=m_{11}$, we have:
\begin{eqnarray}
g_h^{-2}&=&{f^h \over\displaystyle
{ 2\pi (4\pi )^{2\over 3}}} \,
\end{eqnarray}
Having the above information, one can obtain the non-perturbative
superpotential:
\begin{eqnarray}
W=m_{11} (c+
~h~ exp(-{3 \over\displaystyle {b_0 (4\pi )^{5\over 3}}}
(S -  (\sum_{i=1}^3
\alpha_i T_i ))) )\,
\end{eqnarray}

Here, we would like to reemphasize one point. Although
the above results are similar to those in the weakly coupled
string because  the low energy action in
M-theory is similar to that of weakly coupled string,
the difference between the M-theory and the weakly coupled
string is the magnitude of the corrections. In M-theory, they
are relatively very large.
\section{Soft Terms}
We would like to calculate the soft terms that arise in above case. 
We assume the auxiliary components $F^S$ and $F^{T_i}$ of the
moduli superfields $S$ and $T_i$ could get nonvanishing vev and break
SUSY; even though we do not know much about the details of SUSY breaking 
process, we can get the most important information: soft terms.

Apply the standard soft term formulars for the SUGRA
model~\cite{SASW, ABIM, VSKJL}, one can
calculate the soft terms straightforwardly. We do not consider
the bilinear parameter B because it dependent on the particular mechanism
which could generate the $\mu$-term.

First, we write the K\"ahler potential in another
form:
\begin{eqnarray}
K &=& \hat K +
\tilde K_{i\bar i} C_{i a}^* C_i^a ~,~\,
\end{eqnarray}
where i=1, 2, 3 and
\begin{eqnarray}
\hat K = -\ln\,[S+\bar S ]
-\sum_{i=1}^3 \ln\, [T_i +\bar T_i] ~,~ \,
\end{eqnarray}
and
\begin{eqnarray}
\tilde K_{i\bar i}=  ( 2 + {2\over 3} {1\over {S+\bar S}}
(\sum_{j=1}^3 \alpha_j (T_j + \bar T_j)))
{1\over\displaystyle {T_i +\bar T_i}} ~.~\,
\end{eqnarray}
The tree level supergravity scalar potential is
\begin{eqnarray}
V(\phi , \phi^*)&=&(\bar F^{\bar N} K_{\bar N M} F^M - 3 e^G) ~,~ \,
\end{eqnarray}
where G is the K\"ahler function, $K_{\bar N M} \equiv
\partial_{\bar N} \partial_M  K$. The auxiliary fields
$F^M=e^{G/2} K^{M \bar P} G_{\bar P}$ where $K^{M \bar P}$
is the inverse of the $K_{\bar P M}$. Consider $ < C_i^a > =0$
and gravitino mass $m_{3/2}=e^{G/2}$, we can write $F^S$
and $F^{T_i}$ as
the following:
\begin{eqnarray}
F^S &=& {\sqrt 3} sin\theta ~ m_{3/2}
 C (S +\bar S) e^{-i\delta_S} ~,~\,
\end{eqnarray}
\begin{eqnarray}
F^{T_1} &=& {\sqrt 3} cos\theta ~ sin\beta_1 ~
 m_{3/2} C (T_1 +\bar T_1) e^{-i\delta_{T_1}} ~,~\,
\end{eqnarray}
\begin{eqnarray}
F^{T_2} &=& {\sqrt 3} cos\theta ~ cos\beta_1 ~ sin\beta_2 ~
 m_{3/2} C (T_2 +\bar T_2) e^{-i\delta_{T_2}} ~,~\,
\end{eqnarray}
\begin{eqnarray}
F^{T_3} &=& {\sqrt 3} cos\theta ~ cos\beta_1 ~ cos\beta_2 ~
 m_{3/2} C (T_3 +\bar T_3) e^{-i\delta_{T_3}} ~,~\,
\end{eqnarray}
where $C=(1+{V \over\displaystyle {3 m_{3/2}^2}})^{1/2}$
and V is the tree level vacuum energy density.
Therefore, the normalized soft gaugino masses and the un-normalized 
soft scalar masses, and the trilinear parameters are:
\begin{eqnarray}
M_{1/2}&=&{{F^S + \sum_{i=1}^3 \alpha_i F^{T_i}}
\over\displaystyle { (S+ \bar S) +
\sum_{j=1}^3 \alpha_j ( T_j + \bar T_j) }} ~,~\,
\end{eqnarray}
\begin{eqnarray}
m_i^2&=&(m_{3/2}^2 +V) \tilde K_{\bar i i}
\nonumber\\&& - \bar F^{\bar m} (\partial_{\bar m}
\partial_n \tilde K_{\bar i i}-
\partial_{\bar m} \tilde K_{\bar i i}
{1\over\displaystyle \tilde K_{\bar i i}}
\partial_n \tilde K_{\bar i i}) F^n ~,~\,
\end{eqnarray}
\begin{eqnarray}
A_{ a b c }= F^m ( \hat K_m - \sum_{i=1}^3
{1\over \tilde K_{\bar i i}}
\partial_m \tilde K_{\bar i i}) ~,~\,
\end{eqnarray}
where the non-zero $\hat K_m$, $\partial_m \tilde K_{\bar i i} $
and $\partial_{\bar m} \partial_n \tilde K_{\bar i i} $ are
\begin{eqnarray}
\hat K_S ={{-1}\over\displaystyle {S+\bar S}} ~,~ \,
\end{eqnarray}
\begin{eqnarray}
\hat K_{T_i} ={{-1}\over\displaystyle {T_i
+\bar T_i}} ~,~ \,
\end{eqnarray}
\begin{eqnarray}
\partial_S \tilde K_{\bar i i} = -{2\over 3}
{1\over\displaystyle {(S+ \bar S)^2}}
(\sum_{k=1}^3 \alpha_k (T_k + \bar T_k))
{1\over\displaystyle {T_i +\bar T_i}} ~,~ \,
\end{eqnarray}
\begin{eqnarray}
\partial_{\bar S} \partial_S \tilde K_{\bar i i}
&=& {4\over 3}
{1\over\displaystyle {(S+ \bar S)^3}}
\nonumber\\&&
(\sum_{k=1}^3 \alpha_k (T_k + \bar T_k))
{1\over\displaystyle {T_i +\bar T_i}} ~,~ \,
\end{eqnarray}
\begin{eqnarray}
\partial_{\bar T_i} \partial_S \tilde K_{\bar i i}
&=& {2\over 3}
{1\over\displaystyle {(S+ \bar S)^2}}
\nonumber\\&&
(\sum_{j=1, j\not= i}^3 \alpha_k (T_j + \bar T_j))
{1\over\displaystyle {(T_i +\bar T_i)^2}} ~,~ \,
\end{eqnarray}
\begin{eqnarray}
\partial_{\bar T_j} \partial_S \tilde K_{\bar i i}
&=& - {2\over 3}
{{\alpha_j} \over\displaystyle {(S+ \bar S)^2}}
 {1\over\displaystyle {T_i +\bar T_i}} ~,~ \,
\end{eqnarray}
\begin{eqnarray}
\partial_{T_j} \tilde K_{\bar i i}
&=& { 2\over 3}
{{\alpha_j} \over\displaystyle {(S+ \bar S)}}
 {1\over\displaystyle {T_i +\bar T_i}} ~,~ \,
\end{eqnarray}
\begin{eqnarray}
\partial_{\bar T_i} \partial_{T_j} \tilde K_{\bar i i}
&=& - {2\over 3}
{{\alpha_j} \over\displaystyle {(S+ \bar S)}}
 {1\over\displaystyle {(T_i +\bar T_i)^2}} ~,~ \,
\end{eqnarray}
\begin{eqnarray}
\partial_{T_i} \tilde K_{\bar i i}
&=& -( 2+ {2\over 3}
{1\over\displaystyle {(S+ \bar S)}}
\nonumber\\&&
(\sum_{j=1, j\not= i}^3 \alpha_k (T_j + \bar T_j)))
{1\over\displaystyle {(T_i +\bar T_i)^2}} ~,~ \,
\end{eqnarray}
\begin{eqnarray}
\partial_{\bar T_i} \partial_{T_i} \tilde K_{\bar i i}
&=& ( 4+ {4\over 3}
{1\over\displaystyle {(S+ \bar S)}}
\nonumber\\&&
(\sum_{j=1, j\not= i}^3 \alpha_k (T_j + \bar T_j)))
{1\over\displaystyle {(T_i +\bar T_i)^3}} ~,~ \,
\end{eqnarray}
where in all above equations, $j\not= i$.
From the above calculation, we notice that, for the Yukawa coupling,
the universal condition is satisfied. But it seems that we might not 
have the scalar mass universality. However, noticing these conditions
may be reasonable:
$ < T_1 > \sim < T_2> \sim < T_3 >$ and
$\alpha_1 \sim \alpha_2 \sim \alpha_3 $,
we conclude that scalar
mass universality might be approximately satisfied or the violation 
of the scalar mass universality may be very small.
In addition we would like to consider
the dilaton-induced SUSY-breaking, because it is simple, and these
boundary conditions for soft terms may be obtained
for any compactification of M-theory or weakly coupled string.
Moreover, these condintions are universal, gauge group and
flavour independent. But, now the soft masses obtained for
the scalars might be negative, which lead to tachyons unlike
the previous results.
We define $F^S=\sqrt 3 m_{3/2} ( S+\bar S)$ and require $V=0$.
Then, the normalized soft terms are:
\begin{eqnarray}
M_{1/2}&=&{{\sqrt 3 m_{3/2} (S+\bar S)}
\over\displaystyle { (S+ \bar S) +
\sum_{j=1}^3 \alpha_j (T_j +\bar T_j )}} ~,~ \,
\end{eqnarray}
\begin{eqnarray}
m_i^2 & =& m_{3/2}^2 - 3 m_{3/2}^2 (2 f_i -f_i^2) ~,~\,
\end{eqnarray}
\begin{eqnarray}
A_{a b c} &=& \sqrt 3 m_{3/2} ( -1 + f_1 +f_2 +f_3) ~,~\,
\end{eqnarray}
where
\begin{eqnarray}
f_i &=&{2 \over 3} {1\over \tilde K_{\bar i i}}
{1\over (S +\bar S)} {1\over (T_i + \bar T_i)}
(\sum_{j=1}^3 \alpha_j (T_j +\bar T_j )) ~.~\,
\end{eqnarray}
\section{Conclusion}
In the leading order, following the compactification
line from 11-dimension
to 5-dimension, then to 4-dimension, we discuss various
orbifold
compactifications of the M-theory
suggested by Horava and Witten
on $T^6/Z_3$, $T^6/Z_6$, $T^6/Z_{12}$ and the compactification
by keeping singlets under $SU(2)\times U(1)$ symmetry,
 whose Hodge-Betti
numbers are $h_{(1,1)}=9, 5, 3, 2$ and $h_{(2, 1)}=0$ respectively,
then the
compactification on $S^1/Z_2$.
Although the original Lagrangian is order $\kappa^{2/3}$,
we also discuss next-leading order K\"ahler potential,
superpotential, and gauge kinetic function in the $Z_{12}$ case.
In addition, we discuss the non-perturbative superpotential related
to the $S$ and $T_i$. Finally,
we calculate the SUSY breaking soft terms and
find out that the universality of the scalar masses will be
violated, but the violation might be very small.

\section*{Acknowledgments}
I would like to thank V. Barger for reading the manuscript and comments
and thank A. Lukas for helpful e-mail discussions.


\newpage

\begin{thebibliography}{99}
\itemsep 0.5mm
\bibitem{HW} P. Horava and E. Witten, Nucl. Phys. B {\bf475} (1996) 94.
\bibitem{Witten} E. Witten, Nucl. Phys. B {\bf471} (1996) 135.
\bibitem{Horava} P. Horava, Phys. Rev. D {\bf54} (1996) 7561.
\bibitem{AQR} I. Antoniadis and M. Quiros, hep-th/9705037,
hep-th/9707208, hep-th/9709023.
\bibitem{EDCJ} E. Dudas and J. C. Grojean, hep-th/9704177;
 E. Dudas, hep-th/9709043.
\bibitem{LT} Z. Lalak and S. Thomas, hep-th/9707223.
\bibitem{LOW} A. Lukas, B. A. Ovrut and D. Waldram, hep-th/9711197.
\bibitem{CKM} K. Choi, H. B. Kim and C. Munoz, hep-th/9711158
\bibitem{NOY} H. P. Nilles, M. Olechowski and M. Yamguchi,
hep-th/9707143, Phys. Lett. B {\bf415} 415 (1997) 24;
hep-th/9801030.
\bibitem{BD} T. Banks and M. Dine, Nucl. Phys. B {\bf479} (1996) 173 and
hep-th/9609046.
\bibitem{KC} K. Choi, Phys. Rev. D{\bf 56} (1997) 6588.
\bibitem{AQ} I. Antoniadis and M. Quiros, Phys. Lett. B {\bf392}  
(1997) 61.
\bibitem{VK} E. Caceres, V. S. Kaplunovsky and I. M. Mandelberg,
Nucl. Phys. B{\bf 493} (1997) 73.
\bibitem{EFN} J. Ellis, A. E. Faraggi and D. V. Nanopolous,
hep-th/9709049.
\bibitem{MP} E. A. Mirabelli and M. E. Peskin, hep-th/9712214.
\bibitem{LLN} T. Li, J. L. Lopez and D. V. Nanopoulos,
Mod. Phys. Lett. A {\bf12} (1997) 2647.
\bibitem{TIAN} T. Li, J. L. Lopez and D. V. Nanopoulos,
Phys. Rev. D{\bf56} (1997) 2602.
\bibitem{LOD} A. Lukas, B. A. Ovrut and D. Waldram, hep-th/9710208.
\bibitem{DVN} D. V. Nanopolous, hep-th/9711080.
\bibitem{BKL} V. Barger, C. Kao and T. Li, in preparation.
\bibitem{BL} T. Li, in preparation.
\bibitem{BJ} T. Li, in preparation.
\bibitem{JEDVN} J. Ellis, P. Kanti, N. E. Mavromatos
and D. V. Nanopolous, hep-th/9711163.
\bibitem{no-scale} E. Cremmer, S. Ferrara, C. Kounnas, and D. V.  
Nanopoulos,
Phys. Lett. B{\bf133} (1983) 61; J. Ellis, A. Lahanas, D. V. Nanopoulos
and K. Tamvakis, Phys. Lett. B{\bf134} (1984) 429; J. Ellis, C.  
Kounnas, and
D. V. Nanopoulos, Nucl. Phys. B{\bf241} (1984) 406 and B{\bf247}  
(1984) 373.
For a review see A. Lahanas and D. V. Nanopoulos, Phys. Rep.  
{\bf145} (1987) 1.
\bibitem{JLDVN} J. L. Lopez and D. V. Nanopoulos,
Int. J. Mod. Phys. A {\bf11} (1996) 3439.
\bibitem{SFKP} S. Ferrara, C. Kounnas and M. Porrati,
Phys. Lett. B {\bf 181} (1986) 263.
\bibitem{MRNE} M. Dine, R. Rohm, N. Seiberg and E. Witten,
Phys. Lett. B{\bf 156} (1985) 55.
\bibitem{SASW} S. K. Soni and H. A. Weldon,
Phys. Lett. B {\bf 126} (1983) 215.
\bibitem{ABIM} A. Brignole, I. E. Ibanez and C. Munoz,
hep-ph/9707209; Nucl. Phys. B {\bf 422} (1994) 125.
\bibitem{VSKJL} V. S. Kaplunowsky and J. Louis, Phys. Lett.
B{\bf 306} (1993) 269.

\end{thebibliography}
\end{document}